\documentclass[pdflatex,sn-mathphys-num]{sn-jnl}

\usepackage[dvipsnames]{xcolor}
\usepackage{subcaption}
\usepackage{derivative}
\usepackage{wrapfig}
\usepackage{bm}
\usepackage{amsmath,amssymb,amsthm}
\usepackage{mathrsfs}
\usepackage[title]{appendix}
\usepackage{textcomp}
\usepackage{manyfoot}
\usepackage{booktabs}
\usepackage{multirow}
\usepackage{physics}
\usepackage{algorithm}
\usepackage{algorithmicx}
\usepackage{algpseudocode}
\usepackage{listings}

\newcommand{\dbar}{{\mkern3mu\mathchar'26\mkern-12mu d}}  
\theoremstyle{thmstyleone}%
\theoremstyle{thmstyletwo}%

\theoremstyle{thmstylethree}%

\begin{document}

\title[Statistical entropy of quantum systems]{Statistical entropy of quantum systems}


\author[1]{\fnm{Smitarani} \sur{Mishra}}
\author*[1]{\fnm{Shaon} \sur{Sahoo}} \email{shaon@iittp.ac.in}

\affil[1]{\orgdiv{Department of Physics}, \orgname{Indian Institute of Technology Tirupati}, \orgaddress{\street{Yerpedu}, \postcode{517619}, \state{Andhra Pradesh}, \country{India}}}

\abstract{Statistical formulations of thermodynamic entropy, such as those by Boltzmann and Gibbs, were originally developed for classical systems and are well understood in that context. However, the foundational aspects of quantum statistical mechanics remain an area of active debate and are yet to be fully understood. This work is motivated by the need to develop a comprehensive understanding of the statistical measures of thermodynamic entropy in quantum systems - a topic intimately connected to the phenomenon of quantum thermalization. In particular, we investigate the conditions under which the von Neumann entropy can be regarded as a valid statistical measure of thermodynamic entropy in quantum systems. This paper demonstrates that the equivalence between the von Neumann and thermodynamic entropies is not universal, but instead depends on several subtle and often overlooked assumptions. In this context, we also briefly revisit key criticisms of von Neumann entropy — particularly its time-invariance and subadditivity — and argue that these concerns can be meaningfully addressed in the setting of thermodynamic systems. 
To substantiate some arguments and to clarify some issues, we provide suitable numerical results from our analysis of a spin-1/2 system.}




\maketitle

\section{Introduction} \label{sec1}
Since the development of quantum mechanics, there has been an enormous progress in our understanding of how the quantum many-body systems behave. This progress saw some foundational questions being asked and at-least partially answered. One of such questions is on whether and how the statistical mechanics works for the quantum many-body systems \cite{Gogolin-Eisert16, DAlessio-Rigol16, Reimann08}. This gives rise to the important field of quantum statistical mechanics (QSM). The thermodynamic (TH) entropy is a very important state function which helps us understand how a system approaches equilibrium besides playing a crucial role in studying equilibrium properties of a system \cite{Kardar_SPP}. The usefulness of the TH entropy naturally calls for its appropriate quantum counterpart in QSM. 

The TH entropy, as defined by Clausius in the context of second law of thermodynamics, is a state function which takes the maximum value for a system in the equilibrium under fixed macroscopic conditions \cite{Kardar_SPP, Mori18}. For an isolated system, this entropy never decreases when the system undergoes some process. Thus any entropy defined in the similar spirit for a quantum system must be a state function and should increase (or remain constant) when the system undergoes a process.    

Boltzmann, and later Gibbs, showed us how the TH entropy can be calculated from the microscopic details using statistical approaches \cite{Kardar_SPP,Pathria_SM}. For an isolated system, Boltzmann gave a formula (actually Planck first wrote down the explicit expression for the Boltzmann entropy) for the TH entropy in terms of the volume of the relevant phase space (or the number of associated microstates). For the open system, Gibbs on the other hand, using his ensemble theory, provided the formula for the TH entropy in terms of an appropriate probability distribution function in phase space. In Sec. \ref{sec2}, we discuss in detail the conditions under which these statistical measures of the TH entropy are valid and applicable.  

While the Boltzmann and Gibbs entropies (the statistical entropies) are understood and analyzed for the classical systems, their quantum analogues for {\it isolated} and {\it open} quantum systems in equilibrium are often considered intuitively plausible \cite{Goldstein19, Mori18}, although a thorough and convincing justification is still required - an issue we address in Sec. \ref{sec2}. A more subtle and open question concerns the appropriate quantum version of thermodynamic (TH) entropy for a {\it subsystem} of a larger quantum system. Specifically, we aim to understand whether the von Neumann (VN) entropy of a subsystem can be regarded as equivalent to its TH entropy. This question is particularly important because quantum thermalization - and the framework of the Eigenstate Thermalization Hypothesis (ETH) - is fundamentally defined in terms of subsystems or local observables.

We organize our paper in the following way. In the next section (Sec. \ref{sec2}), we review the thermodynamic and the (classical) statistical entropies (Boltzmann's and Gibbs' versions), and discuss the conditions under which the thermodynamic and statistical entropies are equivalent. This is followed by a discussion on the suitable quantum versions of the statistical entropies. Subsequently, we show that these quantum versions of the statistical entropies can be expressed in a basis-independent form, which coincidentally looks exactly like the VN entropy formula. In Sec. \ref{sec3}, we discuss the von Neumann entropy and its interpretations. The discussion also includes some major reservations about the entropy when it comes to representing the thermodynamic entropy, and how to address them meaningfully. Next, in Sec. \ref{sec4}, we start with the VN entropy and analyze if and when it reduces to the TH entropy; this is contrary to what is done in Sec. \ref{sec2}: there we start with the TH entropy and show how in the quantum domain it can be written in a basis-independent form (which, coincidentally, looks like the VN entropy formula). Besides, not only isolated and open systems, we also analyze the important case related to subsystem in Sec. \ref{sec4}. We conclude our work in Sec. \ref{sec6}.   

\section{Thermodynamic and statistical entropies} \label{sec2}
In this section we first briefly review the thermodynamic (TH) entropy and the (classical) statistical (ST) entropies (both Boltzmann's and Gibbs' versions). Next, we try to understand under what conditions these entropies (TH and ST) are equivalent. Subsequently, we discuss how these statistical entropies can be reinterpreted in terms of the quantum variables. At the end of this section, we discuss how the quantum versions of the ST entropies can be expressed in a basis-independent form (which, coincidentally, looks like the VN entropy formula).     

\subsection{Entropy: from Clausius to Boltzmann and Gibbs}
The Clausius theorem or the Clausius inequality states that \cite{Kardar_SPP}, for any cyclic transformation or thermodynamic process, $\oint \frac{\dbar q}{T} \le 0$, where $\dbar q$ is the heat increment received by the system at temperature $T$. Here the equality sign works only for a reversible cycle and this helps us to define a thermodynamic (TH) state function, called entropy. Strictly speaking, this state function (thermodynamic entropy, $S_{TH}$) is well defined only for a thermodynamic system in equilibrium. If a system in equilibrium at temperature $T$ receives $\dbar q$ amount of heat reversibly and subsequently reaches an equilibrium again, then the difference in entropy between these two equilibrium phases would be $d S_{TH} = \frac{\dbar q}{T}$. In contrast, if the process is irreversible, then $d S_{TH} > \frac{\dbar q}{T}$. This shows, for any isolated system undergoing a process ($\dbar q =0$), the change in its entropy is $d S_{TH}\ge 0$.

The Boltzmann's version of statistical entropy is defined by the number of accessible microstates $\Omega$ of a system (with a fixed energy). For a gaseous system, this number is proportional to the phase-space volume available to the system in the given macrostate. Here the statistical entropy is defined as $S_{BN}=k_B \ln {\Omega}$ with $k_B$ being the Boltzmann (BN) constant \cite{Kardar_SPP,Landau_SP,Goldstein19}. By dividing the accessible phase-space as per the values of the well-defined macrovariables, one can in-fact calculate the BN entropy for each such division \cite{Goldstein19, Mori18}. This way one can even make the BN entropy useful in studying a system approaching equilibrium. 

The statistical entropy as per Gibbs (GB), for a given macrostate, is defined as $S_{GB} = -k_B \int \rho (x) \ln {\rho (x)}~ dx$, where $dx$ is an appropriate (symmetrized) volume element of the phase-space and $\rho (x)$ is the density or probability distribution function describing (the ensemble of) the system \cite{Goldstein19}. In principle, one can apply this formula to calculate entropy of a system in a non-equilibrium state (described by a suitable distribution $\rho$). This entropy, interestingly, does not increase (or change) due to the Liouville evolution (the Hamiltonian dynamics). There are many suggestions on how one can resolved this issue and make the GB entropy increase for a system approaching equilibrium \cite{Goldstein19}. We will again come back to this issue in Sec. \ref{sec3c}.

In the following subsection we discuss under what conditions the TH entropy and the statistical entropy (BN or GB entropy) are equivalent. 

\subsection{Conditions of equivalence between thermodynamic and statistical entropies} \label{sec2b}

\paragraph{Case of Boltzmann entropy:} 
To understand the conditions under which the thermodynamic entropy ($S_{TH}$) is equivalent to the Boltzmann entropy ($S_{BN}$), i.e., $S_{TH}\equiv S_{BN}$, one typically considers two systems with a fixed total energy $E$ \cite{Pathria_SM}. 
These systems are initially in thermal equilibrium separately and are then brought into weak thermal contact, allowing energy exchange but not particle transfer. The exchange of heat brings the composite system to a new equilibrium. For thermodynamically large systems, the interaction energy between the systems is assumed negligible.

The analysis relies on the principle of equal \textit{a priori} probabilities, which states that in equilibrium, each accessible microstate is equally likely. For a given energy distribution where the first and the second systems respectively have the energies \(E_1'\) and \(E_2' = E - E_1'\), the total number of microstates for the composite system is given by
\[
\Omega(E_1') = \Omega_1(E_1') \, \Omega_2(E - E_1'),
\]
where \(\Omega_1\) and \(\Omega_2\) are the number of microstates of the two systems at the respective energies.

The total number of microstates of the composite system is
\[
M(E) = \sum_{E_1'} \Omega(E_1'),
\]
where the sum is for all possible energy distributions. In the thermodynamic limit, this sum is sharply peaked at a particular energy partition \(\overline{E}_1\), which maximizes \(\Omega(E_1')\). The system spends most of its time near this most probable energy configuration, and one can approximate 
\(M(E) \approx \Omega(\overline{E}_1) = \Omega_1(\overline{E}_1) \, \Omega_2(E - \overline{E}_1)\).

The above result helps us to derive an effective condition of the equilibrium. It can be obtained by maximizing \(\Omega(E_1')\), or, more conveniently, \(\ln \Omega(E_1')\). Comparing the maximization condition with the thermodynamic identity
\[
\frac{1}{T} = \left( \frac{\partial S_{\mathrm{TH}}}{\partial E} \right)_{N,V},
\]
one finds that the thermodynamic entropy of the first system must be equal to the Boltzmann entropy evaluated at the most probable energy \(\overline{E}_1\), which is given by 
\[
S_{\mathrm{BN}} = k_B \ln \Omega_1(\overline{E}_1).
\] 

From the above discussion, we conclude that the TH entropy and the BN entropy of an isolated system are equivalent ($S_{TH}\equiv S_{BN}$) if the system is large (so that the temperature $T$ can be defined and $M(E)$ can be approximated by $\Omega(\overline{E}_1)$) and the system is in thermal equilibrium (so that the principle of equal \textit{a priori} probabilities becomes applicable).  

\paragraph{Case of Gibbs entropy:}

To establish the equivalence between the thermodynamic (TH) entropy and the Gibbs (GB) entropy, \( S_{\mathrm{TH}} \equiv S_{\mathrm{GB}} \), one first determines the equilibrium probability distribution function \( \rho_{\mathrm{eq}}(x) \). This is done by considering a setup similar to the one described earlier: a target system in weak thermal contact with a much larger system acting as a reservoir \cite{Pathria_SM}. Assuming the interaction energy between the systems is negligible, and both systems are in thermal equilibrium, the equilibrium distribution for a microstate \( x \) of the target system is given by
\[
\rho_{\mathrm{eq}}(x) = \frac{1}{Q} e^{-\beta H_1(x)},
\]
where \( H_1(x) \) is the energy of the target system when it is in the microstate \( x \), and \( Q \) is the partition function,
\[
Q = \int e^{-\beta H_1(x)} \, dx.
\]

This distribution yields the average (internal) energy \( U \) of the system in terms of the partition function \( Q \). Comparing the resulting expression for \( U \) with the thermodynamic identity
\[
U = \left[ \frac{\partial (A/T)}{\partial (1/T)} \right]_{N, V},
\]
one can identify the thermodynamic (Helmholtz) free energy \( A \) in terms of the partition function $Q$. This link between thermodynamics and statistical mechanics allows us to show that the thermodynamic entropy in equilibrium is equal to the Gibbs entropy, given by
\[
S_{\mathrm{GB}} = -k_B \int \rho_{\mathrm{eq}}(x) \ln \rho_{\mathrm{eq}}(x) \, dx.
\]
This equivalence holds under the assumptions mentioned above: large system size, negligible interaction energy, and thermal equilibrium.

\subsection{Quantum version of statistical entropy} \label{sec2c}
In the preceding subsection, we discuss the conditions under which the thermodynamic entropy ($S_{TH}$) is equivalent to the statistical entropy ($S_{BN}$ or $S_{GB}$) for classical systems. In this subsection, we explore how these conditions translate to the quantum domain. Following this discussion, we introduce the plausible quantum analogues of the classical statistical entropies.
  
\paragraph{Equivalence conditions in quantum domain:}

While the quantum analogue of a classical statistical entropy ($S_{BN}$ or $S_{GB}$) may seem intuitively apparent, rigorously establishing its validity is more subtle. In particular, it is not straightforward to determine whether the conditions discussed in the previous subsection — originally derived for classical systems — hold in the quantum setting, or whether they must be modified. For instance, the principle of equal \textit{a priori} probabilities, which underlies much of classical statistical mechanics, requires careful reconsideration in the quantum context.

In a classical system, the time evolution carries the system through different microstates — each point along a constant-energy trajectory is considered a microstate and is statistically equally probable. But what constitutes a microstate in a quantum system? A natural first guess might be to associate the microstates with the energy eigenstates of the system’s Hamiltonian. This would then allow us to consider all eigenstates within a narrow energy window and assume equal probability among them, mimicking the classical principle.

However, this approach faces a conceptual difficulty: a quantum system in a pure energy eigenstate does not evolve into other eigenstates under unitary time evolution. This static nature of energy eigenstates seems incompatible with the dynamical perspective of statistical ensembles. In the following, we discuss how this issue can be addressed and how a consistent statistical framework can still be formulated for quantum systems.

If we consider an isolated large quantum system with size $N$ (the `size' can be interpreted as the number of particles in or constituents of the system), the average gap between any two consecutive eigenvalues will be exponentially small with the system size. We note that the (average) density of states is proportional to the Hilbert space dimension ($D$) and the average gap ($\delta$) is inversely proportional to the density of states. Since $D\propto e^{cN}$, we have $\delta \propto e^{-cN}$, where $c$ is a positive constant. Thus, if $N$ is  large enough, it will be practically impossible to prepare the quantum system in a pure state. 
In realistic settings, therefore, we are limited to describing the system as occupying a set of closely spaced eigenstates with some probability distribution. In equilibrium, and in the absence of any bias toward specific states within a narrow energy window, it is reasonable to invoke the principle of equal \textit{a priori} probabilities and assume a uniform probability distribution across these states.

There are other ways to justify the principle of equal probabilities in a quantum system. One argument relies on the system's possible entanglement with its environment due to weak interactions in the past. Even if the system is currently isolated, such prior interactions can leave it in an entangled state, effectively rendering it a mixed state. As a result, the system may occupy any of the microstates (i.e., energy eigenstates) within a certain energy range, and it is natural to assign a uniform probability distribution over these states.

Another alternative justification comes from perturbation theory. Let \( H_0 \) be the main Hamiltonian of the non-interacting system, and \( H' \) a weak interaction term, so that the total Hamiltonian is \( H = H_0 + H' \). If the eigenstates of \( H_0 \) are regarded as microstates, then the interaction \( H' \) drives transitions among them over time. When \( H' \) is sufficiently weak, all states within a narrow energy band become accessible, and assigning equal probabilities to them is justified -recovering the framework of statistical mechanics.

Once we have agreement on the quantum version of the principle of equal \textit{a priori} probabilities, it is not difficult to come up with the quantum versions of the statistical entropies; we discuss these below. 

\paragraph{Quantum version of Boltzmann entropy:}

To define the quantum analogue of the Boltzmann (qBN) entropy, we consider a system prepared such that its energy lies within a narrow energy shell \((E, E + \Delta E]\), where \(\Delta E\) is arbitrarily small on macroscopic scales but large enough microscopically to include many energy eigenstates \cite{Mori18}. Let \(\{E_n\}\) denote the set of eigenvalues of the system's Hamiltonian, with \(\{\ket{n}\}\) as the corresponding eigenstates. According to the equal probability principle, the system is equally likely to be in any state \(\ket{n}\) such that \(E_n \in (E, E + \Delta E]\). Drawing on our earlier discussion of the classical Boltzmann entropy, the quantum version, \(S_{qBN}\), can then be written as:
\begin{equation}
    S_{BN} \to S_{qBN} = k_B \ln D,
    \label{s_qbn}
\end{equation}
where \(D\) is the number of eigenstates (microstates) with energies in the specified shell.

It is worth noting that the energy shell can be further partitioned into macrostates (i.e., orthogonal subspaces), each containing microstates consistent with a specific set of macroscopic properties \cite{Goldstein19, Mori18}. Among these, the macrostate with the largest dimension dominates in the thermodynamic limit. For a large system, the dimension of this equilibrium subspace becomes essentially equal to \(D\), justifying the use of the full shell in the entropy definition.

\paragraph{Quantum version of Gibbs entropy:}
To formulate the quantum version of the Gibbs (qGB) entropy, we replace the continuous classical energy variable with the discrete energy eigenvalues of the system’s Hamiltonian when computing the partition function. We assume that the system interacts weakly with a thermal reservoir (or did so in the past), allowing us to describe its dynamics effectively using its own Hamiltonian. Let \(\{E_n\}\) be the set of energy eigenvalues of the (open) system. In analogy with the classical Gibbs distribution, we assign to each energy level \(E_n\) a probability
\[
p_n = \frac{1}{Q} e^{-\beta E_n},
\]
where \(Q = \sum_n e^{-\beta E_n}\) is the partition function and \(\beta\) is the inverse temperature. The quantum analogue of the Gibbs entropy, $S_{qGB}$, is then given by
\begin{equation}
    S_{GB} \to S_{qGB} = -k_B \sum_n p_n \ln p_n.
    \label{s_qgb}
\end{equation}



\subsection{A basis-independent representation of quantum statistical entropy} \label{sec2d}
In this subsection, we will see how one can express the quantum versions of the statistical entropies in a basis-independent form. 

For an isolated system, described by a microcanonical (mc) ensemble, we get the following from Eq. \ref{s_qbn}: $S_{qBN} = -k_B \ln D = -k_B \text{Tr} (\rho_{mc} \ln \rho_{mc})$, where $\rho_{mc} = \frac{1}{D} I$ with $D$ being the dimension of the relevant subspace of the Hilbert space of the system and $I$ is the identity matrix. 

 Similarly, for an open system, we get the following from the Eq. \ref{s_qgb}, $S_{qGB} = -k_B\sum_n p_n\ln {p_n} = - k_B \text{Tr} (\rho_{\beta} \ln \rho_{\beta})$, where $\rho_{\beta} = \frac{e^{-\beta H}}{\text{Tr} (e^{-\beta H})}$. Here $H$ and $\beta$ are, respectively, the Hamiltonian and the inverse temperature of the open system. 

We note that, in both the cases, the basis-independent expression coincidentally looks exactly like the VN entropy expression, as discussed later (Eq. \ref{s_vn}). It may be mentioned here that, while deriving the expression of the thermodynamic entropy of a subsystem (weakly connected to rest of the system), Landau also came up with the same basis-independent representation of the quantum statistical entropy \cite{Landau_SP}.

\subsection{Summary of the current section and some remarks}

In this section (\ref{sec2}), we demonstrated that the thermodynamic entropy \( S_{\text{TH}} \) coincides with the Boltzmann entropy \( S_{\text{BN}} \) for isolated classical systems, and with the Gibbs entropy \( S_{\text{GB}} \) for open classical system - provided certain standard conditions are met. We then introduced their quantum counterparts, denoted \( S_{\text{qBN}} \) and \( S_{\text{qGB}} \), respectively. Notably, both quantum statistical entropies can be recast in a basis-independent form that is formally identical to the von Neumann (VN) entropy. This correspondence suggests that, under appropriate conditions - such as large system size and thermal equilibrium - the VN entropy serves as a quantum analogue of the thermodynamic entropy for both isolated and open quantum systems.

While this section has focused on isolated and open systems, we have yet to address the case of a subsystem embedded within a larger quantum system. This scenario will be explored in detail in Sec.~\ref{sec4}. There, we revisit the equivalence question for all three cases - isolated systems, open systems, and subsystems - by first explicitly computing the VN entropy in each setting and then comparing it with the corresponding thermodynamic entropy.

\section{von Neumann Entropy} \label{sec3}
While reformulating some of the major aspects of statistical mechanics so that it can be used in studying quantum systems, von Neumann provided a formula for entropy of a quantum system in a mixed state \cite{Neumann_MFQM}. If the density matrix $\rho$ represents the mixed state of a system (not to be confused with the probability distribution function denoted by the same symbol), the entropy of the system, according to von Neumann, is 
\begin{equation}
    S_{VN} = - k_B \text{Tr} (\rho \ln \rho).
    \label{s_vn}
\end{equation}
This formula attracted both the applause and criticism alike \cite{Shenker99,Hemmo06,Henderson03,Deville13,Strasberg21,Sheridan20}. In the following subsections, we first discuss two different perspectives to understand how one can arrive at the formula and what it actually quantifies. Next, we highlight some of the major criticisms on the applicability of the formula and explain our views that many of such criticisms can be avoided or can be addressed convincingly.

\subsection{Background}
In many situations, we may lack precise knowledge of the exact pure state in which a quantum system resides. To address this, von Neumann introduced the concept of the \textit{density matrix} (\(\rho\)) as a statistical tool for describing such systems \cite{Neumann_MFQM}. The density matrix represents a \textit{statistical mixture} of pure states and is expressed as  
\begin{equation}
    \rho = \sum_{i} p_i \ket{\psi_i}\bra{\psi_i},
    \label{rho_mixed}
\end{equation}
where each pure state \(\ket{\psi_i}\) occurs with probability \(p_i\), and \(\sum_i p_i = 1\).  

To compute the entropy associated with a system in the mixed state \(\rho\), von Neumann introduced the idea of an \textit{ensemble} analogous to the Gibbs ensemble in classical statistical mechanics. In this ensemble, the state \(\rho\) is represented by \(\mathcal{N} p_i\) copies of the pure state \(\ket{\psi_i}\), where \(\mathcal{N}\) (a large number) denotes the total number of copies in the ensemble.

While following the Gibbsian way, one immediately faces problems in defining entropy of the system in the state $\rho$. First of all, the decomposition in Eq. \ref{rho_mixed} is not unique and, then, the pure states $\{\ket{\psi_i}\}$ in a decomposition may not be in general distinguishable (as they are not in general orthonormal). Through thermodynamic considerations, von Neumann proposed a formula (as given in Eq. \ref{s_vn}) to quantify the entropy of a system in state $\rho$. We will not go through his arguments here; rather, we will discuss two alternative ways to understand the VN entropy. Before that, we briefly discuss below the two main quantum processes and their effects on the VN entropy.

As discussed by von Neumann, we can consider two basic processes that a system may undergo: (1) time evaluation of the system, and (2) measurement on the system. If $\rho(0)$ is the density matrix at time $t=0$ then the density matrix at the $t$ is given by $\rho(t) = U(t) \rho(0) U^{\dag}(t)$, where $U(t)$ is the time evaluation operator. It is not difficult to see that the VN entropy is invariant under this time evaluation process, i.e., $S_{VN}(t) = S_{VN}(0)$. von Neumann showed that a measurement, in general, irreversibly changes the density matrix to another density matrix $\rho \to \rho'$, where 
\begin{equation}
    \rho' = \sum_n \bra{\phi_n}\rho\ket{\phi_n} P_n.
    \label{rho_qm}
\end{equation}
We note that the set of states $\{\ket{\phi_n}\}$ are the eigenkets of the observable for which one is performing the measurement and the projection operator $P_n = \ket{n}\bra{n}$. The subtleties in dealing with degeneracy can be found in von Neumann's work \cite{Neumann_MFQM}. He showed that the entropy of a system, in general, increases (or remains constant but never decreases) due to the measurement process, i.e., $S_{VN}(\rho')\ge S_{VN}(\rho)$. If we identify the first process (time evaluation) and the second process (measurement) with the thermodynamic reversible and irreversible processes, respectively, then these results are consistent with the thermodynamic laws. 

In the following, we now discuss two different ways one can intuitively understand the entropy formula (Eq. \ref{s_vn}).

\subsection{Interpretations of VN entropy}
\paragraph{First interpretation:}
When a system is in a pure state (considered a microstate), then the entropy of the system, be it \( S_{qBN} \) or \( S_{VN} \), is zero.   
However, when the system is in a mixed state - a statistical mixture of pure states described by the density matrix \( \rho \) - the entropy must quantify the degree of ``mixedness'' or statistical impurity in \( \rho \), analogous to how the Boltzmann entropy $S_{BN}$ reflects the uncertainty in locating a system within a particular microstate.

Looking at the decomposition in Eq.~\ref{rho_mixed}, one might be tempted to define the entropy as the Shannon (SN) entropy of the probability distribution \( \{p_i\} \), i.e.,
\[
S_{SN}(\{p_i\}) = -k_B \sum_i p_i \ln p_i.
\]
However, this approach fails because the decomposition of a mixed state into pure states is not unique - different ensembles can give rise to the same density matrix. Consequently, entropy defined this way would not be well-defined.

One way to resolve this ambiguity is to consider all possible ensemble decompositions of \( \rho \) and select the one that minimizes the Shannon entropy \( S_{SN}(\{p_i\}) \). This approach guarantees a unique value of entropy for a given \( \rho \), independent of the specific decomposition. In fact, it can be shown that the von Neumann entropy of \( \rho \) corresponds precisely to this minimum value \cite{Jaynes57,Facchi21}:
\begin{equation}
S_{VN}(\rho)=inf \left\{ S_{SN}(\{p_i\}) : \rho = \sum_{i}p_i \ket{\psi_i}\bra{\psi_i} \right\}.
\label{svn_mindec}
\end{equation} 

\paragraph{Second interpretation:}
Another way to understand the VN entropy is as follows. 
If we interpret a set of orthonormal states as a collection of distinguishable microstates, we can define an entropy based on the probabilities associated with these states. Let \(\{\ket{\phi_n}\}\) be a set of orthonormal states. Then the Shannon (SN) entropy associated with the probability distribution \(w_n = \bra{\phi_n} \rho \ket{\phi_n}\) is given by  
\[
S_{SN}(\{w_n\}) = -k_B \sum_n w_n \ln w_n.
\]  
However, this entropy is not uniquely defined, as it depends on the particular choice of the orthonormal basis \(\{\ket{\phi_n}\}\) used to represent the microstates.

To address this issue, consider constructing a new density matrix \(\rho'\) as  
\[
\rho' = \sum_n w_n \ket{\phi_n}\bra{\phi_n},
\]  
where again \(w_n = \bra{\phi_n} \rho \ket{\phi_n}\). Each \(\ket{\phi_n}\) is now an eigenvector of \(\rho'\) with eigenvalue \(w_n\), and the matrix \(\rho'\) takes a form analogous to that in Eq.~\ref{rho_qm}. Importantly, this new density matrix \(\rho'\) can be viewed as the post-measurement state when a projective measurement is made in the basis \(\{\ket{\phi_n}\}\).

As originally demonstrated by von Neumann \cite{Neumann_MFQM}, such a projective measurement leads to an increase in entropy:
\(
- \mathrm{Tr}(\rho' \ln \rho') \geq - \mathrm{Tr}(\rho \ln \rho).
\)
Therefore, the von Neumann entropy represents the minimum possible Shannon entropy over all orthonormal decompositions, and this minimum is achieved when the orthonormal basis \(\{\ket{\phi_n}\}\) consists of the eigenkets of the given density matrix \(\rho\).
Accordingly,
\begin{equation}
S_{VN}(\rho)=inf \bigg\{ S_{SN}(\{w_i\}) : w_i=\bra{\phi_i}\rho\ket{\phi_i} \bigg\},
\label{svn_minqm}
\end{equation}
 where it is understood that $\bra{\phi_i}\ket{\phi_j}=\delta_{ij}$.

\subsection{Addressing objections against VN entropy} \label{sec3c}     
Over the century since von Neumann’s seminal work, the entropy bearing his name has found widespread applications across quantum physics and information theory. However, its interpretation — particularly as a candidate for thermodynamic (TH) entropy — has not been without criticism. In what follows, we address two of the most significant objections commonly raised against interpreting the von Neumann (VN) entropy as the thermodynamic entropy.

\paragraph{Time-invariance of VN entropy:} 
The first major objection concerns the fact that the VN entropy of an isolated quantum system remains constant over time as the system evolves unitarily. Specifically, under unitary evolution \(\rho(0) \to \rho(t) = U(t)\rho(0)U^\dagger(t)\), the entropy satisfies \(S_{VN}(t) = S_{VN}(0)\). In contrast, the thermodynamic (TH) entropy of an isolated system is expected to increase as the system approaches thermal equilibrium. This naturally raises the question: does the VN entropy fail to describe entropy in non-equilibrium situations?

It is interesting to note here that this time-invariance of the VN entropy has a close analogue in classical statistical mechanics: the Gibbs (GB) entropy is also conserved under Hamiltonian evolution. This is because the shape of the phase-space probability distribution remains invariant during such evolution \cite{Goldstein19,Frigg22}. 
Various approaches have been proposed to resolve the issue of time-invariance of the GB entropy; detailed discussions may be found elsewhere \cite{Goldstein19,Frigg22}.

Here, we also emphasize that the issue of the time-invariance arises primarily when we consider systems out of equilibrium and seek to assign an entropy value during a dynamical process. However, in thermal equilibrium, under broad and general conditions discussed earlier, the VN entropy can be considered equivalent to the TH entropy for isolated and open quantum systems (see Sec.~\ref{sec2d}). Consequently, the VN entropy remains a valid tool for determining the (thermodynamic) entropy differences between equilibrium states, thereby capturing the entropy change associated with thermodynamic processes.

In addressing the time-invariance issue of VN entropy, two broad approaches have been considered. The first suggests modifying or generalizing the VN entropy formula so that the resulting entropy increases over time and eventually saturates in equilibrium. A well-known example of this approach is the \textit{observational entropy}, which has gained considerable attention in recent years \cite{Strasberg21,Safranek21,Nagasawa24}. The second approach does not question the VN entropy formula itself; rather, it proposes that the conventional time evolution of the density matrix \(\rho\), governed by the von Neumann-Liouville equation, may not fully capture the correct dynamics, especially for systems undergoing thermalization. In this work, we adopt the second viewpoint, and our reasons for doing so are as follows.

First, like TH entropy, the VN entropy is a \textit{state function} — it depends solely on the system’s state \(\rho\) and is independent of any observer-dependent choices. Furthermore, the VN entropy is known to be equivalent to the TH entropy in open or isolated system. 
In fact, starting from the VN entropy formula (Eq. \ref{s_vn}) and invoking the maximum entropy principle along with standard constraints, one can derive the equilibrium Gibbs distribution \cite{Sakurai_MQM}. These observations strongly support the adequacy of the VN entropy formula for studying quantum thermodynamics.

The real challenge, then, lies in understanding how the density matrix evolves from an initial non-equilibrium state to a final state representing thermal equilibrium. In this work, we do not delve into the details of such irreversible dynamics. However, we briefly mention that, while describing irreversible evolution for an isolated quantum system remains a complex problem, the dynamics of open quantum systems are better understood and can be effectively described by the Gorini-Kossakowski-Sudarshan-Lindblad (GKSL) master equations \cite{Yoshida24}.

Finally, we also note in passing that, in certain special cases, the two aforementioned approaches — modifying the entropy formula vs.\ modifying the evolution equation — may actually coincide. For example, in the case of a single coarse-graining procedure, the observational entropy turns out to be exactly equal to the VN entropy computed using the coarse-grained density matrix \cite{Safranek21,Wehrl78}.

\paragraph{Subadditivity of VN entropy:}
The second major objection to the VN entropy is that it satisfies the \textit{subadditivity} property, whereas the TH entropy is expected to be \textit{additive} - at least for a large, uniform system in which the interaction energy between its subsystems can be neglected. The subadditivity property states that for a bipartite quantum system \( AB \), described by a joint state \( \rho_{AB} \), with subsystems \( A \) and \( B \) having reduced states \( \rho_A = \mathrm{Tr}_B(\rho_{AB}) \) and \( \rho_B = \mathrm{Tr}_A(\rho_{AB}) \), the VN entropy satisfies the inequality \cite{Lieb70}:
\begin{equation} \label{subadd}
S_{VN}(\rho_{AB}) \leq S_{VN}(\rho_A) + S_{VN}(\rho_B).
\end{equation}
This has an important consequence: if the composite system is in a pure state (so that \( S_{VN}(\rho_{AB}) = 0 \)), the reduced states \( \rho_A \) and \( \rho_B \) can still be mixed, and their individual VN entropies can be strictly positive---unless the overall state is a product state. This behavior contradicts the \textit{additivity} property typically expected of the TH entropy, which asserts that the entropy of a whole system should be the sum of the entropies of its (non-interacting or weakly interacting) parts.
We can circumvent this apparent contradiction in the following way. 

To examine whether the VN entropy satisfies the additivity property, we begin by noting that quantum thermalization is typically defined with respect to a subsystem of a much larger isolated quantum system \cite{Deutsch91,Srednicki94,Gogolin-Eisert16,Mori18,choi24,DAlessio-Rigol16}. The focus is on whether the subsystem thermalizes, with the remainder of the system effectively acting as a heat bath. Therefore, it is more appropriate to investigate the additivity of entropy at the subsystem level rather than for the entire system. This mirrors the classical approach, where the additivity of thermodynamic (Gibbs) entropy is studied for an open system in equilibrium with a reservoir. In that setting, additivity means that the entropy of the open system is simply the sum of the entropies of its parts (here, the entropy of the bath does not play any role). 

\begin{figure}
    \centering
    \includegraphics[width=8.0cm, height=6.0cm]{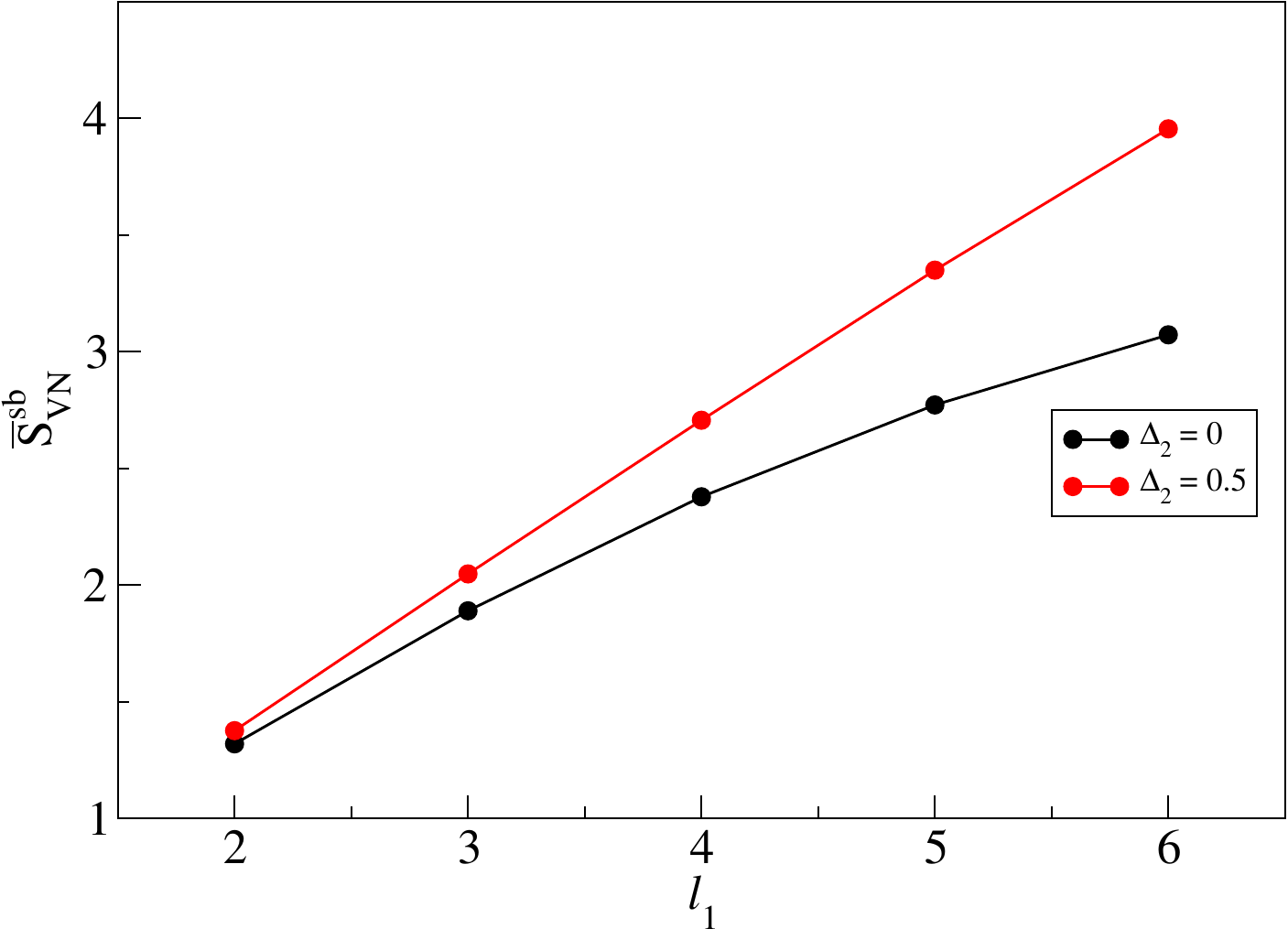}
    \caption{The average VN entropy, $\overline{S}^{sb}_{VN}$, is plotted as a function of the subsystem size $l_1$. Here total system size is $N=16$.}
    \label{svn_size}
\end{figure}

Building on the above discussion, we now examine whether the von Neumann (VN) entropy of a subsystem satisfies the additivity property. It is well established that for highly excited states - particularly those near the middle of the spectrum - the VN entropy exhibits a volume law, scaling linearly with the subsystem size \cite{Molter14,Alba09,Moudgalya18,Bianchi22}.
We numerically demonstrate this scaling for our spin model in Fig. \ref{svn_size}. The detailed discussion on this spin model can be found in Sec. \ref{sec5}.
The linearity of the scaling implies that the entropies of the parts of the subsystem add up to the entropy of the total subsystem. Thus, when properly interpreted, the VN entropy does respect the additivity property in thermalized subsystems. 

It may be also noted here that, for large system size ($N$), the linearity of the $\overline{S}^{sb}_{VN}$ vs. $l_1$ plot is seen for small subsystem size $l_1$. While for the nonintegrable system, the plot is seen to retain the linearity even for larger value of $l_1$, the plot slowly deviates from linearity as $l_1$ is increased beyond a small value for the integrable system. This result also goes well with our conclusion (see Sec. \ref{sec4c}) that, for integrable systems, the VN entropy is in general not equal to the TH entropy, when the full system is in a typical pure state.

We finish this subsection by concluding that the major criticisms against the VN entropy either can be avoided or can be addressed convincingly if issues are analyzed appropriately. In this regard, we concur with A. Peres when he says that {\it there should be no doubt that von Neumann's entropy is equivalent to the entropy of classical thermodynamics} \cite{Peres_QTCM}. 

\section{VN entropy of large quantum systems in equilibrium} \label{sec4}
We explained in Sec. \ref{sec2} that for the isolated and open systems, the quantum versions of, respectively, the Boltzmann and Gibbs entropies can be put in a basis-independent form which coincidentally looks exactly like the von Neumann (VN) formula for entropy. This indirectly implies that, in equilibrium and under certain very general conditions, the VN entropy is equivalent to the thermodynamic entropy (or quantum version of Boltzmann or Gibbs entropy) for isolated or open quantum system. 

In this section, we show this equivalence in a different way. We first consider the state ($\rho$) of a system (isolated or open) in equilibrium, then calculate the VN entropy and show that it is equal to the corresponding quantum statistical entropy ($S_{qBN}$ or $S_{qGB}$). Besides, we also show here that, even for a subsystem of a large quantum system, the VN entropy is equivalent to the TH entropy. This is demonstrated by showing that the average VN entropy corresponding to a narrow energy window is proportional to logarithm of the density of states (DOS) of the full system. 

In the following, we discuss the issue of equivalence separately for the isolated, open and subsystem (of a much larger system). 

\subsection{Isolated quantum system} \label{sec4a}
We consider an isolated quantum system in thermal equilibrium. As discussed in Sec. \ref{sec2c}, a large isolated quantum system will be in a mixed state, for one or multiple reasons. An isolated system in thermal equilibrium is appropriately described by a microcanonical (mc) ensemble.   
Assume that we prepare the system in such a way that its energy lies in the narrow energy shell $(E,E+\Delta E]$, where $\Delta E$ is arbitrary but macroscopically small and microscopically large \cite{Mori18}. Let $\{\ket{n}\}$ be the set of eigenkets of the Hamiltonian of the system, and $E_n$ be the eigenvalue corresponding to $\ket{n}$. The density matrix which describes the isolated system is given by,
\begin{equation}
    \rho_{mc}=\sum_{\substack{n\\E_n\in \left( E,E+\Delta E \right] }} p_n \ket{n}\bra{n},
    \label{rho_fulsys}
\end{equation}
where $p_n$ is the probability to find the system in state $\ket{n}$. 
Since $\Delta E$ is macroscopically small, all the accessible microstates (here, the eigenkets) will essentially have almost the same energy. Application of the principle of equal \textit{a priori} probabilities suggests that $p_n=\frac 1D$, where $D$ is the number of eigenstates associated with the energy shell (see also Sec. \ref{sec2c}).
Since $S_{VN} = -k_B \text{Tr}(\rho_{mc} \ln \rho_{mc})$, we need the eigenvalues of $\rho_{mc}$ to efficiently compute $S_{VN}$. It is clear from Eq. \ref{rho_fulsys} that $\ket{n}$ is an eigenket of $\rho_{mc}$ with eigenvalue $\frac 1D$, i.e., $\rho_{mc}\ket{n}=\frac{1}{D}\ket{n}$ for any energy eigenket from the shell. This implies that
\begin{equation}
    S_{VN} = -k_B \text{Tr}(\rho_{mc} \ln \rho_{mc}) = k_B\ln D.
    \label{svn_fulsys}
\end{equation}
But this is the quantum version of the Boltzmann entropy $S_{qBN}$ (the representative of the thermodynamic entropy). Thus, 
\begin{equation}
    S_{VN} = k_B\ln D = S_{qBN} \equiv S_{TH}.
    \label{sth_fulsys}
\end{equation}

This shows that, for large isolated quantum system in equilibrium, the VN entropy is equivalent to the TH entropy. 

\subsection{Weakly interacting open quantum system}
Here we consider an open system that is in weak contact with the thermal bath. Although the system exchanges heat with the bath, we assume that the interaction with the bath is so weak that the energy spectrum of the system is more or less unaffected by the interaction, and the dynamics of the quantum system can still be described (at-least effectively) by its Hamiltonian $H$.   

As discussed in Sec. \ref{sec2c} and \ref{sec2d}, the density matrix that appropriately describes the open system is $\rho_{\beta} = \frac{e^{-\beta H}}{Tr (e^{-\beta H})}$, where $\beta$ is the inverse temperature. The VN entropy of the system is given by
\begin{equation}
    S_{VN} = -k_B \text{Tr}(\rho_{\beta} \ln \rho_{\beta}) = -k_B\sum_n p_n\ln {p_n},
    \label{svn_opsys}
\end{equation}
where $p_n=\frac{1}{Q}e^{-\beta E_n}$ with $Q$ being the partition function, $Q=\sum_ne^{-\beta E_n}$. Here, $\{E_n\}$ is the set of eigenvalues of the system Hamiltonian $H$.

The right-hand side of Eq. \ref{svn_opsys} is just the quantum version of Gibbs entropy $S_{qGB}$ (the representative of the thermodynamic entropy here). Thus,
\begin{equation}
    S_{VN} = -k_B\sum_n p_n\ln {p_n} = S_{qGB} \equiv S_{TH}.
    \label{sth_opsys}
\end{equation}

This shows that, for large open quantum system
in equilibrium, the VN entropy is equivalent to the TH entropy.

\begin{figure}[h]
\centering
\includegraphics[width=0.95\textwidth]{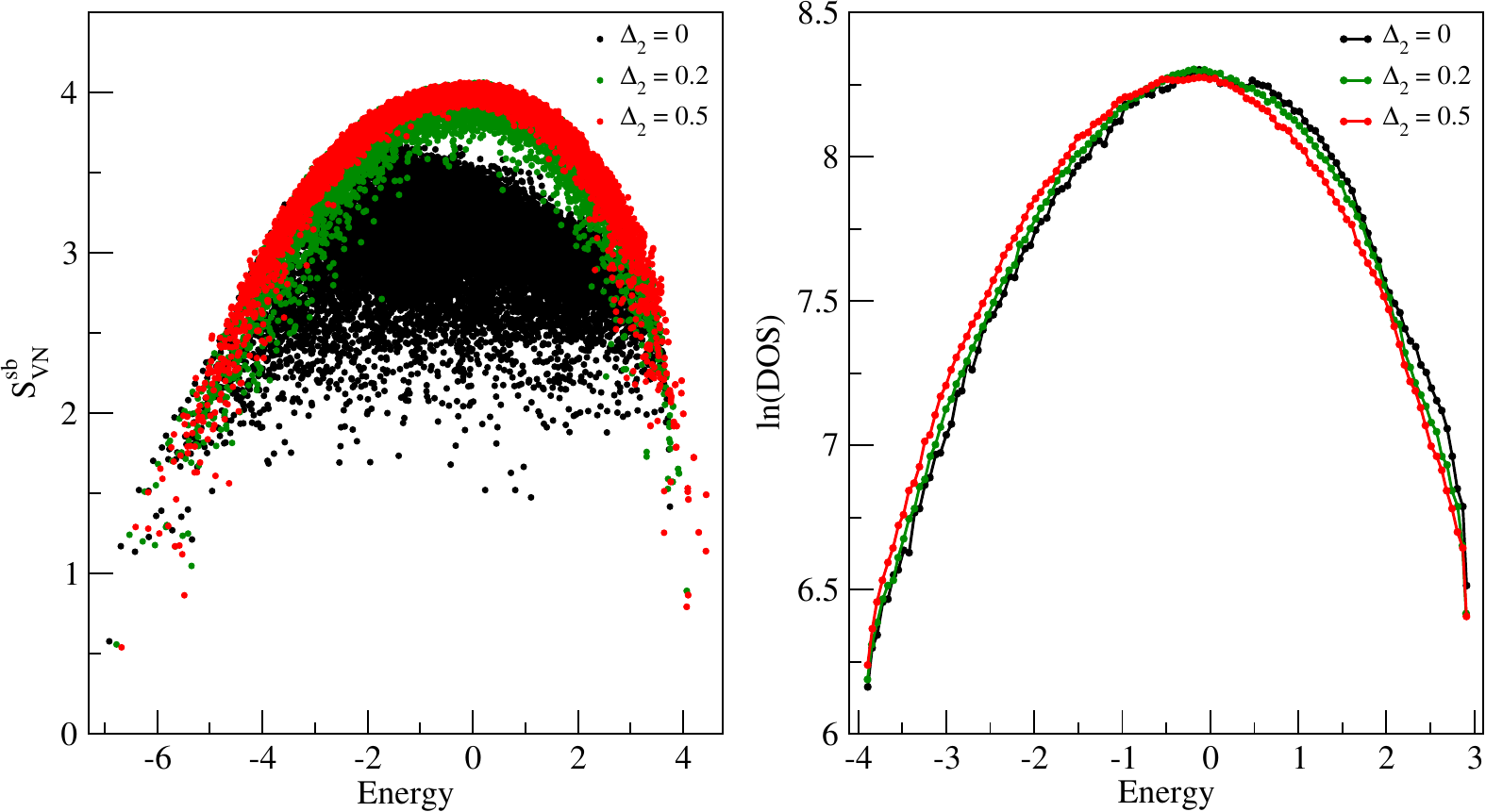}
\caption{The subsystem VN entropy, $S^{sb}_{VN}(\rho_{sb})$, for the individual eigenkets and $\ln (DOS)$ are plotted against the energy of the full system. Here total system size is 16 and subsystem size is 6.}
\label{ee_lndos_eng}
\end{figure}

\begin{figure}
    \centering
    \includegraphics[width=0.7\textwidth]{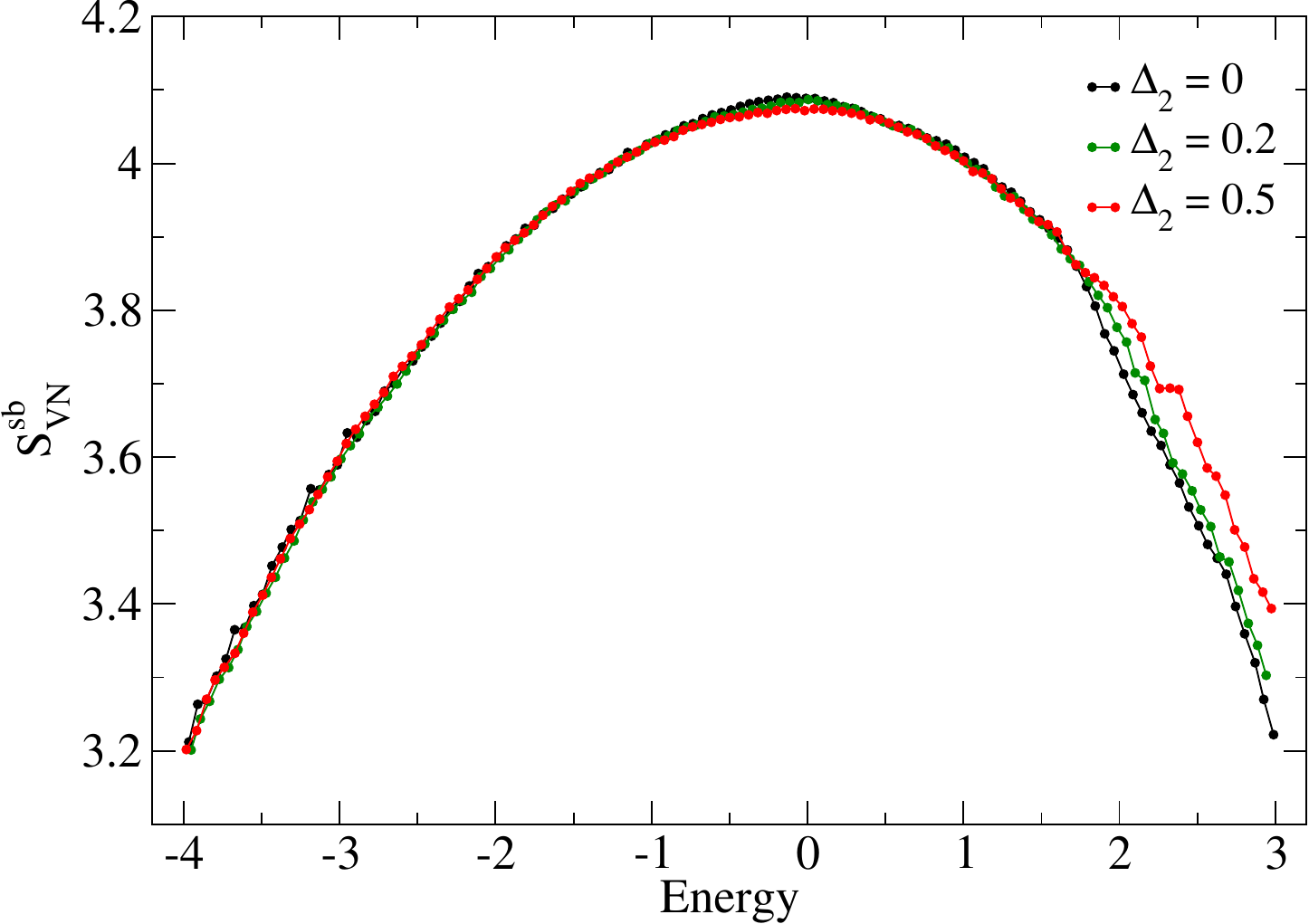}
    \caption{The subsystem VN entropy, $S^{sb}_{VN}(\overline{\rho}_{sb})$, when the full system is in the microcanonical density matrix $\rho_{mc}$. Here, the system (subsystem) size is 16 (6).}
    \label{svn_sb_ardm}
\end{figure}

\begin{figure}
    \centering
    \includegraphics[width=7.5cm, height=6cm]{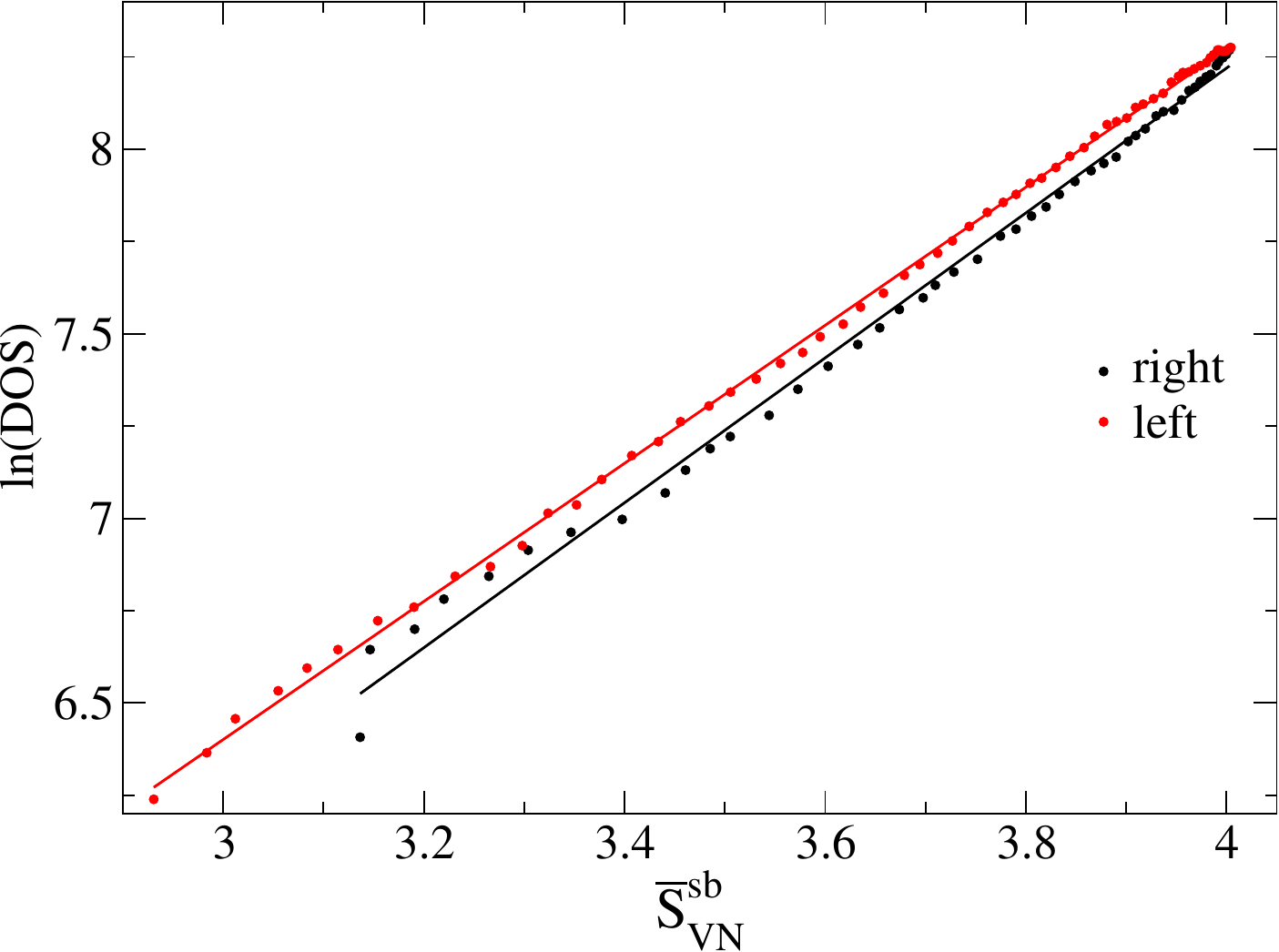}
    \caption{The average VN entropy, $\overline{S}^{sb}_{VN}(\rho_{sb})$, of subsystem is plotted against $\ln (DOS)$ across the energy spectrum (separately for the left and the right halves of the spectrum). Here, the system (subsystem) size is 16 (6) and $\Delta_2=0.5$.}
    \label{lndos_svn}
\end{figure}

\subsection{Subsystem of a quantum system}\label{sec4c}
We separately discuss here the question of equivalence between the subsystem TH and VN entropies when the full system is in a mixed state (described by a microcanonical ensemble) or in a pure state (both {\it typical} and {\it atypical}). 

\paragraph{Full system is in mixed state:}
If the full system is in a mixed state \( \rho_{mc} \), described by the microcanonical ensemble as in Eq. \ref{rho_fulsys}, it can then be shown easily that the reduced state of a small subsystem typically follows a Gibbs distribution \( \rho_\beta \)~\cite{Popescu06, Goldstein06}. Specifically, if \( \overline{\rho}_{sb} = \text{Tr}_B(\rho_{mc}) \) denotes the reduced density matrix of the subsystem (obtained by tracing out the degrees of freedom of the bath \( B \), i.e., the rest of the system), then
\[
\overline{\rho}_{sb} \approx \rho_\beta = \frac{e^{-\beta H}}{Q}, \quad \text{with} \quad Q = \text{Tr}_{sb}(e^{-\beta H}),
\]
where \( H \) is the Hamiltonian of the subsystem and $\beta$ is the inverse temperature (determined by the energy scale of the full system).  

It is now evident that the VN entropy of the subsystem, i.e., $S_{VN}^{sb}$ ($=S_{VN}(\overline{\rho}_{sb})$; see Eq. \ref{s_vn}), will be the same as the quantum version of the Gibbs entropy, i.e., $S_{qGB}$. As discussed in Sec. \ref{sec2c}, under certain general conditions, $S_{qGB}$ is equivalent to the thermodynamic entropy ($S_{TH}$). Thus we see that, for a subsystem, it VN entropy is equivalent to TH entropy if the full system is in a mixed state. 

There is also an alternative way of understanding this. Since $\overline{\rho}_{sb}$ is directly derived from $\rho_{mc}$, which is the uniform mixture of the energy eigenkets from a narrow energy shell, it is expected that $\overline{\rho}_{sb}$ would behave very similarly for both the integrable and nonintegrable systems. This is also evident from Fig. \ref{svn_sb_ardm}, where $S_{VN}(\overline{\rho}_{sb})$ is plotted for both integrable and nonintegrable systems, and found them to be quite similar. 

Now, for a given system (irrespective of integrability), we consider \[S_{VN}(\overline{\rho}_{sb})\approx k_B\ln{d_{sb}},\] 
where $d_{sb}$ is a parameter, interpreted to be the effective Hilbert space dimension of the subsystem when the full system is on a given energy scale. Following \cite{Mishra2025}, we can find the scaling: 
\[d_{sb} =D_{sb}^{\gamma},\] 
where $D_{sb}$ is the dimension of the subsystem's actual Hilbert space and $\gamma=\ln(d_E)/\ln(D)$. In the scaling factor $\gamma$, $d_E$ is the dimension of the subspace of the full Hilbert space, associated with the given energy scale of the full system, and $D$ is the dimension of the full Hilbert space. We note that the value of $\gamma$ is expected to be the same for the integrable and nonintegrable cases. This can be contrasted with the case where the full system is in a pure state (see the next paragraph); $\gamma$ in such a situation is expected to be different for non-integrable (chaotic) and integrable cases.
From this scaling of $d_{sb}$, it is straightforward to see that the VN entropy of the subsystem, $k_B\ln d_{sb}$, is proportional to the logarithm of the density of states (DOS), producing a linear plot as in Fig. \ref{lndos_svn}. This result again confirms that the VN entropy of a subsystem is equivalent to its TH entropy (or qBN entropy) if the full system is in a mixed state.    

\paragraph{Full system is in a pure state:}

If the full system is in a pure state, for the overwhelming majority of pure states, the subsystem can be shown to be described by a Gibbsian distribution $\rho_{\beta}$ \cite{Popescu06,Goldstein06}. This important result is known as {\it canonical typicality} or {\it general canonical principle}. If $\rho_{sb}$ is the reduced density matrix of the subsystem if the full system is in a (given) pure state, then $\rho_{sb}\approx \rho_{\beta}$ for most of the pure states. Now, arguing as before (see previous paragraph on the case of mixed state), we can conclude that the VN entropy $S_{VN}(\rho_{sb})$ for a subsystem is equivalent to its TH entropy (or qGB entropy) when full system is in a pure state, for most of the cases.

This approach leaves open the possibility of pure states (of the full system) for which a subsystem is not described by a canonical Gibbsian distribution. There is an alternative way to show the equivalence between the entropies when the full system is in pure state; this approach also reveals when the VN entropy of subsystem is not equal to the corresponding TH entropy.  

In this alternative approach, the VN entropy $S_{VN}(\rho_{sb})$ is calculated for each eigenket (of the full system) from a narrow energy window. As discussed in \cite{Mishra2025}, one can show that the average VN entropy $\overline{S}_{VN}$ corresponding to an energy scale is proportional to the logarithm of the density of states (DOS), as can be seen in Fig. \ref{lndos_svn}. The proportionality constant $\gamma$ is equal to $\ln(d_E)/\ln(D)$ for nonintegrable (chaotic) system and less than that value for integrable systems. This result implies that the VN entropy of a subsystem is equal to its TH entropy (qBN entropy) when a chaotic system is in a pure typical state, and the VN entropy is less than the TH entropy when an integrable system is in a pure typical state.  

\section{A spin model for numerical analysis} \label{sec5}
For our numerical results, we take the following one-dimensional (with open boundary condition) Heisenberg spin-1/2 chain with the next-nearest neighbor interactions:
\begin{equation}
    H=\sum_{i=1}^{N-1}\vec{S}_i\cdot \vec{S}_{i+1}+\Delta_2\sum_{i=1}^{N-2} S^z_i S^z_{i+2}.
    \label{spn_ham}
\end{equation}
In our calculations, we take the size of the system $N=16$. This model system is integrable when $\Delta_2=0$ and is non-integrable when $\Delta_2> 0$ \cite{Steinigeweg13}. 

For a given $\Delta_2$ value, using the exact diagonalization (ED) method, we solve the Hamiltonian matrix and find all the eigenvalues and eigenvectors corresponding to the sector $S_z=0$ ($S_z$ is the $z$-component of the total spin of the system). From the set of energy eigenvalues, one can then obtain the density of states (DOS) profile, as shown in Fig. \ref{ee_lndos_eng} (b). To calculate the VN entropy corresponding to individual eigenkets, as shown in Fig. \ref{ee_lndos_eng} (a), we first calculate reduced density matrix $\rho_{sb}$ corresponding to each eigenket. This helps us to get the VN entropy $S_{VN}$ for individual eigenkets using Eq. \ref{s_vn}. Corresponding to each energy shell, the average of this VN entropy $\overline{S}_{VN}$ is plotted against the logarithm of DOS in Fig. \ref{lndos_svn}. Now, to calculate the VN entropy of the subsystem when the full system is in mixed state (as shown in Fig. \ref{svn_sb_ardm}), we first average the reduced density matrices corresponding to individual eigenkets associated with a narrow energy shell. This gives us $\overline{\rho}_{sb}$; this is then used to calculate $S_{VN}(\overline{\rho}_{sb})$ using Eq. \ref{s_vn}.

It may be worth mentioning here that, the result that the subsystem's average VN entropy is proportional to system's $\ln (DOS)$, is earlier reported in \cite{Sahoo12} in a different context. We also note that, elsewhere \cite{Sharma13}, the TH entropy of a subsystem is defined as what we call here the VN entropy of a subsystem, $S^{sb}_{VN}(\overline{\rho}_{sb})$, when the full system is in the mixed state $\rho_{mc}$.    

\section{Conclusion}\label{sec6}
Many foundational issues of quantum statistical mechanics are analyzed in this paper and presented in a unified way. 
In reviewing major objections to the von Neumann (VN) entropy as a counterpart to the thermodynamic (TH) entropy, we show that concerns regarding its time invariance and subadditivity can be convincingly addressed. We identify general conditions under which the VN and TH entropies coincide in three scenarios: isolated systems, open systems, and subsystems. For isolated and open systems in equilibrium, the equivalence is relatively straightforward to establish. However, for a subsystem embedded in a much larger system, the equivalence is more subtle. We explain that if the full system is in a mixed state, the subsystem’s VN entropy generally equals its TH entropy. For a pure state of a nonintegrable full system, the two entropies typically agree, whereas differences may arise when the full system is integrable. 
This understanding of the entropy of a subsystem can be important, as quantum thermalization is typically formulated in terms of a subsystem embedded within a much larger system.

\bmhead{Acknowledgements}
SS thanks S. Aravinda for useful discussions. 

\bibliography{manuscript}       

\end{document}